\documentclass[%
 aip,
 sd,%
 amsmath,amssymb,
reprint,%
]{revtex4-1}

\usepackage{graphicx}
\usepackage{dcolumn}
\usepackage{bm}
\usepackage{amsmath}
\begin{document}


\title[Josephson junction based thermometer and its possible application in bolometry]{Josephson junction based thermometer and its application in bolometry }

\author{T. Faivre}
\email{timothe.faivre@aalto.fi}
\affiliation{Low Temperature Laboratory (OVLL), Aalto University, POB 13500, FI-00076 AALTO, Finland
}%
\author{D. Golubev}%
\affiliation{Low Temperature Laboratory (OVLL), Aalto University, POB 13500, FI-00076 AALTO, Finland
}%
 \affiliation{Karlsruhe Institute of Technology (KIT), Institute of Nanotechnology, 76021 Karlsruhe, Germany}
\author{J.P. Pekola}
\affiliation{Low Temperature Laboratory (OVLL), Aalto University, POB 13500, FI-00076 AALTO, Finland
}%

\date{\today}

\begin{abstract}
We propose a new type of a transition edge sensor based on an Al/AlOx/Ti/AlOx/Al superconductor - insulator - superconductor - insulator - superconductor
(SIS'IS) structure. It exhibits sharp dependence of zero bias resistance on temperature of the titanium absorber in the vicinity  
of its superconducting critical temperature. We demonstrate temperature sensitivity of the device  to be $2\ \mu \text{K}/\sqrt{\text{Hz}}$. 
Noise Equivalent Power (NEP) of the device, limited by the amplifier noise, 
is estimated to be $4\times 10^{-17}$ W/$\sqrt{\rm Hz}$ at 313 mK. 
The tunnel junctions between superconducting leads should help to overcome the size limitation imposed by proximity effect 
in conventional transition edge sensors, without sacrificing the sensitivity. Besides that, the input resistance
of the device can be tuned in a wide range.
\end{abstract}

\pacs{74.25.fc, 74.50.+r}
\maketitle
\section{Introduction}
Josephson junctions are used in a variety of applications ranging from radio-frequency 
mixers to magnetic resonance imaging of human tissues in medicine. 
Here we propose a novel Josephson junction based  radiation detector. More specifically, 
we consider a SIS'IS structure with aluminum leads and titanium island in the middle.
We demonstrate that the resistance of this structure is very sensitive
to the temperature of the island in the vicinity of titanium transition
temperature, $T_c^{\rm Ti}$, at which Josephson critical currents of the junctions have strong temperature dependence. 

The sharpness of the superconducting phase transition is often
utilized for metrological purposes. For example, transition temperatures of 
several metals are used as the reference points in approximating the official international temperature scale ITS-90\cite{its}.
It is therefore not surprising, that transition edge sensors (TES) made use of this sharpness to detect radiation\cite{Irwin, richards, cabrera, karasik_review}. 
Conventional TES contains
a superconducting absorber well connected to the superconducting leads 
and kept at a temperature close to the critical 
one. The heating of the absorber by incoming radiation results 
in a change of its resistance. 
TES sensors have excellent sensitivity to THz radiation, extremely low noise levels
and low input impedance, which is necessary for good coupling with 
sensitive SQUID detectors.  
One of the realisations of a TES employs a titanium absorber
coupled to niobium leads\cite{0026-1394-46-4-S28} and resembles
our structure. The sensitivity of TES technique is limited by such factors as the proximity
induced superconductivity in the absorber, and the escape of excited quasiparticles 
into the leads\cite{karasik_review}. 
Detectors with normal island in the middle, i.e. SINIS detectors, have also been proposed
for bolometric applications \cite{Kuzmin,Kuzmin2}. Their sensitivity is comparable to that of TES,
but their input impedance is typically much higher. 
Sensors based on an SNS Josephson junction have been recently proposed as well\cite{Jonas}.
The sensitivity of an SIS'IS sensor, which we propose here, is similar to that of a TES 
and SINIS sensors, and in analogy to those SIS'IS sensor can be used at terahertz frequencies.  
However, it has several important differences, which may
be favorable for certain applications. First, the presence of tunnel junctions suppresses the proximity effect\cite{dubos}.
Second, our device should work at low bias, which potentially makes it less noisy and less dissipative.
Third, zero bias resistance of the device varies over many decades, typically from
hundreds of kOhms down to zero, which allows one to tune the input impedance and choose,
for example, the value close to 50 $\Omega$. This choice may reduce the device sensitivity to
some extent, but improves the coupling to high frequency electronics. 

\section{Sample geometry \& fabrication }

An SEM image of our structure is shown in Fig. 1a.
The junctions have been fabricated with electron beam lithography and shadow angle deposition. 
The aluminum metal was oxidized in situ, on top of the first deposited layer. A buffer layer of aluminum  of few nanometers  
has been then deposited before electron beam evaporation of titanium in order to ensure high quality of the junctions. 
High evaporation rate of 40-50 \AA/s has been used, which allowed us to achieve a critical temperature of titanium 
close to 300 mK reproducibly. 
The normal state specific resistance of the junctions was typically 400 $\Omega \ \mu \text{m}^2$. 
All measurements have been performed in a dilution refrigerator. Room temperature noise has been filtered out 
with lossy coaxial lines thermalized at the mixing chamber temperature. 
SIS'IS sample has been measured in a four-probe scheme. 
Normal state resistance of the structure has been measured at 4 K and equals $2R_n=1240$ $\Omega$. The gap of aluminum
was found to be $\Delta_{\rm Al}=203.4$ $\mu$eV, the titanium gap at zero temperature was $\Delta_{\rm Ti}=44.6$ $\mu$eV,
the corresponding critical temperature was $T_c^{\rm Ti}\approx 315$ mK, 
capacitance of a single junction was estimated to be $C=40$ fF,
and the volume of the absorber was ${\cal V}=0.27$ $\mu$m$^3$. 
\section{Theorical model}
\subsection{Heat balance equation}
Before discussing the results of our measurements in detail, we briefly
introduce the theoretical model of the device.
The bath temperature is kept at temperature $T_{ph}$ close to, but below $T_c^{\rm Ti}$. 
Weak energy input $\dot{Q}_{ext}$ from external sources leads to the elevation of the electron temperature 
of the absorber, $T$, which should be found from linearized heat balance equation 
\begin{equation}
c(T){\cal V}\dot T=\dot{Q}_{ext} - G_{th}(T) (T-T_{ph}).
\end{equation}
Here we have introduced the 
specific heat of the absorber $c(T)\approx c(T_c^{\rm Ti})= \gamma T_c^{\rm Ti} +L$, where $\gamma=315$ J/$\text{m}^3 $/$\text{K}^2 $ is Sommerfeld constant of titanium
and $L=0.43\gamma T_c^{\rm Ti}$ is 
the contribution due to the latent heat of the superconducting transition. The 
 thermal conductance $G_{th}$ is the sum of three contributions,
\begin{equation}
G_{th}= 5 \Sigma {\cal V} T^4 + 2 G_{th}^{jct}+ G_{\nu},
\label{Gth}
\end{equation} 
where the first term is due to electron phonon coupling in the absorber\cite{apl74_3020}
with $\Sigma=1.56\times 10^{-9}$ W/$\mu$m$^3$K$^5$ being the material constant of titanium, the second one describes
heat escape through the junctions and
the last term, $G_{\nu}$, accounts for heat leakage to the electromagnetic environment\cite{Jonas,Meschke2006187}, which we will neglect below for simplicity.
Thermal conductance of a Josephson junction at bias currents below the critical one and at temperature close to 
$T_c^{\rm Ti}$ may be roughly estimated as (see e.g. Ref. \onlinecite{PhysRevB.87.094522} for more details)
\begin{eqnarray}
G_{th}^{jct} \approx \frac{\sqrt{2\pi} k_B\Delta_{\rm Al}^{5/2}}{e^2R_n(k_BT)^{3/2}} e^{-{\Delta_{\rm Al}}/{k_BT}}
+ \beta \frac{(k_BT)^{3/2}V}{\sqrt{\Delta_{\rm Al}}eR_n}.
\label{Gjct}
\end{eqnarray}
Here the first term is the low temperature expansion of thermal conductivity of normal metal - superconductor tunnel junction at zero bias voltage,
and the second term is the contribution of phase slip or phase diffusion processes. 
The dimensionless parameter $\beta\sim 1$ depends on the shape of the resistive branch of the junction I-V curve.  
We find that in our device the thermal conductance is dominated by the electron-phonon contribution, which equals 
 $5 \Sigma {\cal V} T^4 \approx 16$ pW/K.  The contribution of the junctions is lower and equals $G_{th}^{jct} \approx 1 $ pW/K.
The relaxation time of the device may be estimated as
$\tau = c(T){\cal V}/G_{th}\approx 2$ $\mu$s.  
\subsection{Transport in SIS'IS tunnel junction}
Next we turn to the I-V curve
of our device. We assume that the junctions have
identical parameters and define density of states of the leads in the Dynes form  
$
\nu_j(E)= \left| \Re \left( (E+i \eta_j )/\sqrt{(E+i \eta_j)^2-\Delta_{j}^2}  \right) \right|,
$ 
where the index $j$ refers to either titanium or aluminum. The quasiparticle current through the junction 
reads\cite{PhysRevLett.101.077004}
\begin{eqnarray}
I_{qp} &=& \frac{1}{eR_n}\int dE\, \nu_{\rm Al}(E-eV/2)\,\nu_{\rm Ti}(E)
\nonumber\\ && \times\,
\big[ f_{\rm Al}(E-eV/2,T)-f_{\rm Ti}(E,T)\big],
\label{Iqp}
\end{eqnarray}
where $V$ is the total voltage drop across the two junctions and $f_j(E,T_j)$ are the quasiparticle distribution
functions in the leads and in the absorber. By fitting the I-V curves at $T_{bath}>T_c^{\rm Ti}$ to Eq. (\ref{Iqp}),
we have found $\eta_{\rm Al}=3.4 \times 10^{-3} \Delta_{\rm Al}$.
Below the critical temperature of titanium the Josephson junction is described by the usual equation
\begin{eqnarray}
C\frac{\hbar\ddot\varphi}{2e}+\frac{1}{R_{qp}(T)}\frac{\hbar\dot\varphi}{2e}+I_c\sin\varphi = I +\xi(t),
\label{dynamics}
\end{eqnarray}
where $C$ is the junction capacitance, $R_{qp}(T)$ is zero bias quasiparticle resistance of a single junction
derived from Eq. (\ref{Iqp}),
$\varphi$ is the Josephson phase, $I$ is the bias current, 
\begin{equation}
I_c= \frac{1}{ e R_n}\int_{\Delta_{\rm Ti}(T)}^{\Delta_{\rm Al}} dE 
\frac{\Delta_{\rm Ti}(T) \Delta_{\rm Al}[1-2f_{\rm Ti}(E,T)] }{\sqrt{E^2-\Delta_{\rm Ti}^2(T)}\sqrt{\Delta_{\rm Al}^2-E^2}}
\label{eq:ic}
\end{equation}
is the critical current of a single junction\cite{PhysRevLett.10.486}, $\xi(t)$ is the noise
with the spectral density $S_\xi=2k_BT^*/R_{qp}(T)$, where $T^*$ is the noise temperature of the environment.
Since $I_c$ is small in the vicinity of $T_c^{\rm Ti}$, the junction
should be overdamped or only slightly underdamped. Hence one can approximately 
put $C=0$ in Eq. (\ref{dynamics}) while discussing zero bias resistance. In this way we arrive at the 
expression for zero bias resistance of our device in the form\cite{PhysRevLett.22.1364}  
\begin{equation}
R_0(T)=\frac{2R_{qp}(T)}{I^2_{0}\left({\hbar I_c(T)}/{2e k_B T^*}\right)} ,
\label{eqn:zbslope}
\end{equation} 
where $I_0$ is the modified Bessel function of the first kind. Equation (\ref{eqn:zbslope}) describes both the growth 
of the resistance with decreasing temperature at $T_{bath}>T_c^{\rm Ti}$ as well as its sharp drop 
at temperatures below $T_c^{\rm Ti}$. We emphasize that sharp dependence of zero bias resistance on the absorber temperature
$T$ comes from the temperature dependence of the critical current (\ref{eq:ic}) and the
superconducting gap $\Delta_{\rm Ti}(T)$. The environment temperature $T^*$ may differ from $T$.
From Eqs. (\ref{eq:ic},\ref{eqn:zbslope}) one can estimate maximum theoretical responsivity
of the device
\begin{eqnarray}
\frac{\partial R_0}{\partial T}\bigg|_{\max} \approx
0.065 \frac{R_Q^2}{R_n^2}\frac{R_{qp}(T_c^{\rm Ti})}{T_c^{\rm Ti}}
\left(\ln\frac{\Delta_{\rm Al}}{k_BT_c^{\rm Ti}}+0.82\right)^2, 
\label{Smax}
\end{eqnarray}
where $R_Q=h/e^2$ is the resistance quantum.    
For the parameters of our sample we find $\partial R_0/\partial T|_{\max}\approx 3.7\times 10^8$ $\Omega$/K.
In reality the transition is always broadened due to, for example, sample inhomogeneity, fluctuations of the superconducting
order parameter etc., and the responsivity stays below this value.
 
\section{Experimental results}
\subsection{Sample characterisation}
\begin{figure}[h]
\centering
\includegraphics[width=.9\linewidth, height=.9\linewidth]{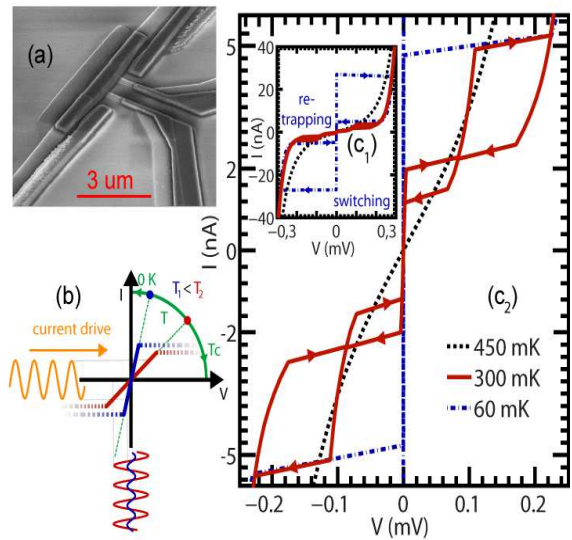}
\caption[SEM picture of the sample and its IV characteristics for few bath temperature]
{(a) An SEM picture of a typical sample. Dark grey corresponds to the titanium layer and the lighter grey lines to aluminium. 
The larger junctions are used as a thermometer and the smaller ones can be used 
for heating or cooling the absorber (not presented here). 
(b)  An artistic view of the zero bias slope measurement. 
An AC bias current is applied to the structure, voltage response provides the information about the absorber temperature. 
(c1) IV characteristics of the structure for three different bath temperatures 
above, close to and far below $T_c^{\rm Ti}\approx 315$mK. 
(c2) Close-up to distinguish the consecutive switching of the junctions. 
The origin of the hysteresis is thermal, the arrows are indicating the direction of the switching.}
\label{fig:iv}
\end{figure}

\begin{figure}
\centering
\includegraphics[width=1\linewidth, height=1\linewidth]{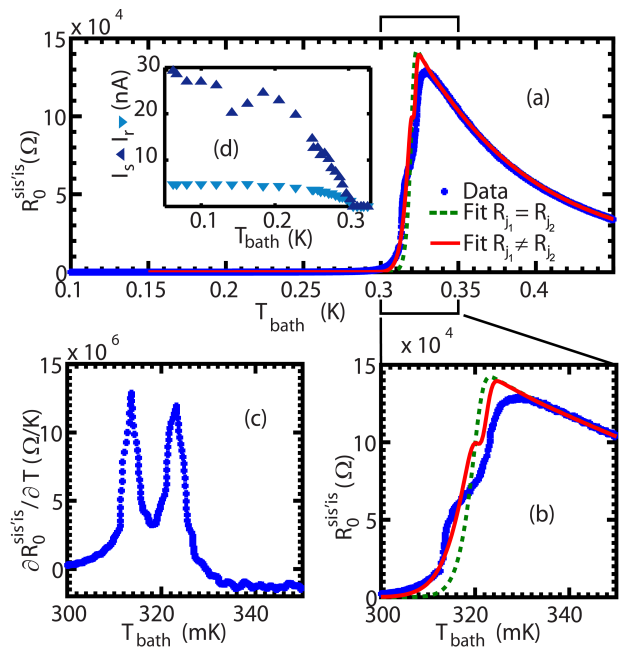}
\caption[zero bias resistance, responsivity and critical current temperature dependence]
{(a) Zero bias resistance versus bath temperature (blue points) and fits to Eq. (\ref{eqn:zbslope}) (dashed green). 
The inset shows switching and retrapping currents emphasizing the non-hysteretic behavior near $T_c$. 
(b) Shape of the transition curve close to $T_C^{\rm Ti}$. Unequal  junctions (red curve) can explain the small step in the middle of the experimental curve, as one them switches before the other.  
(c) Responsivity $\partial R_0/\partial T$. }
\label{fig:resp}
\end{figure}

Figure 1c shows the measured I-V curves of our device. Due to heating effects 
they are hysteretic. The switching to the resistive branch  
occurs at currents much lower than the theoretical value of $I_c$ (\ref{eq:ic}).
We measure zero bias resistance of the structure in a current biased scheme (Fig. \ref{fig:iv}b). 
An AC current drive is applied, the voltage response is measured in a four-probe configuration using a lock-in detector, 
and its amplitude provides the measure of the electronic temperature $T$ of the titanium absorber. 
The phonon temperature, assumed to be equal to the bath temperature, 
is measured with a calibrated ruthenium oxide thermometer attached to the sample stage. 
The amplitude of the probe current is kept fixed during a temperature sweep. 
Close to $T_c^{\rm Ti}$, where the critical current becomes very small, 
the probe current becomes sufficient by large to make the junctions switch to the resistive state. Since in addition, one of the junctions has a larger resistance than the other one, reducing its critical current, one should be able to observe the switching of each junction independently as the temperature increases. This is explaining the broadening and double step shape of the transition curve in Fig. \ref{fig:resp} (a \& b).
The responsivity of the thermometer, ${\partial R_0}/{\partial T}$, is extracted by numerical derivation. 
The maximum in the responsivity is achieved close to the critical temperature (Fig. \ref{fig:resp}c)
and equals $1.3\times 10^7$ $\Omega$/K, which is approximately 30 times  lower than the theoretical expectation (\ref{Smax}). 
The logarithmic derivative  of the resistance, $\alpha =d \ln R / d\ln T$, which characterizes the sharpness of the transition of our sample, $\alpha \approx 200 $, is comparable to the usual Ti-TES sensors \cite{0026-1394-46-4-S28,irwin_apl}.
The half-width of the transition, $\delta T_{\max}$, determined from Fig. \ref{fig:resp}c  roughly equals 2 mK.
Thus the sensor may absorb up to $G_{th}\delta T_{\max}\approx 34$ fW of optical power load without being saturated.  
\subsection{Sensitivity measurements}
\begin{figure}
\centering
\includegraphics[width=1\linewidth, height=1\linewidth]{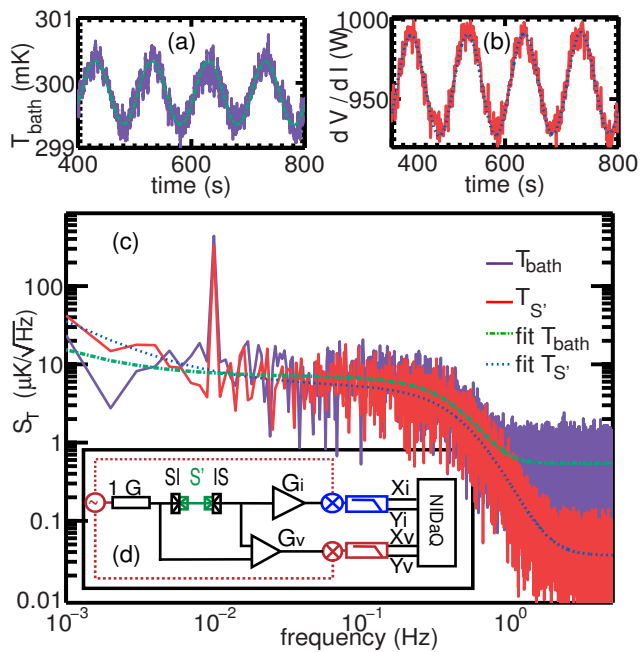}
\caption[responsivity,sensitivity and fft's]{ (a\&b) AC temperature drive measured by the calibrated ruthenium oxide thermometer (a) and the SIS'IS thermometer (b). (c) Fast Fourier transform of the two signals, allowing us to extract the noise level of the thermometers. (d) Schematic setup were the lock-in amplifiers are depicted as mixers followed by a low pass filter.}
\label{fig:fft}
\end{figure}

\begin{figure}[h]
\centering
\includegraphics[width=1\linewidth, height=.9\linewidth]{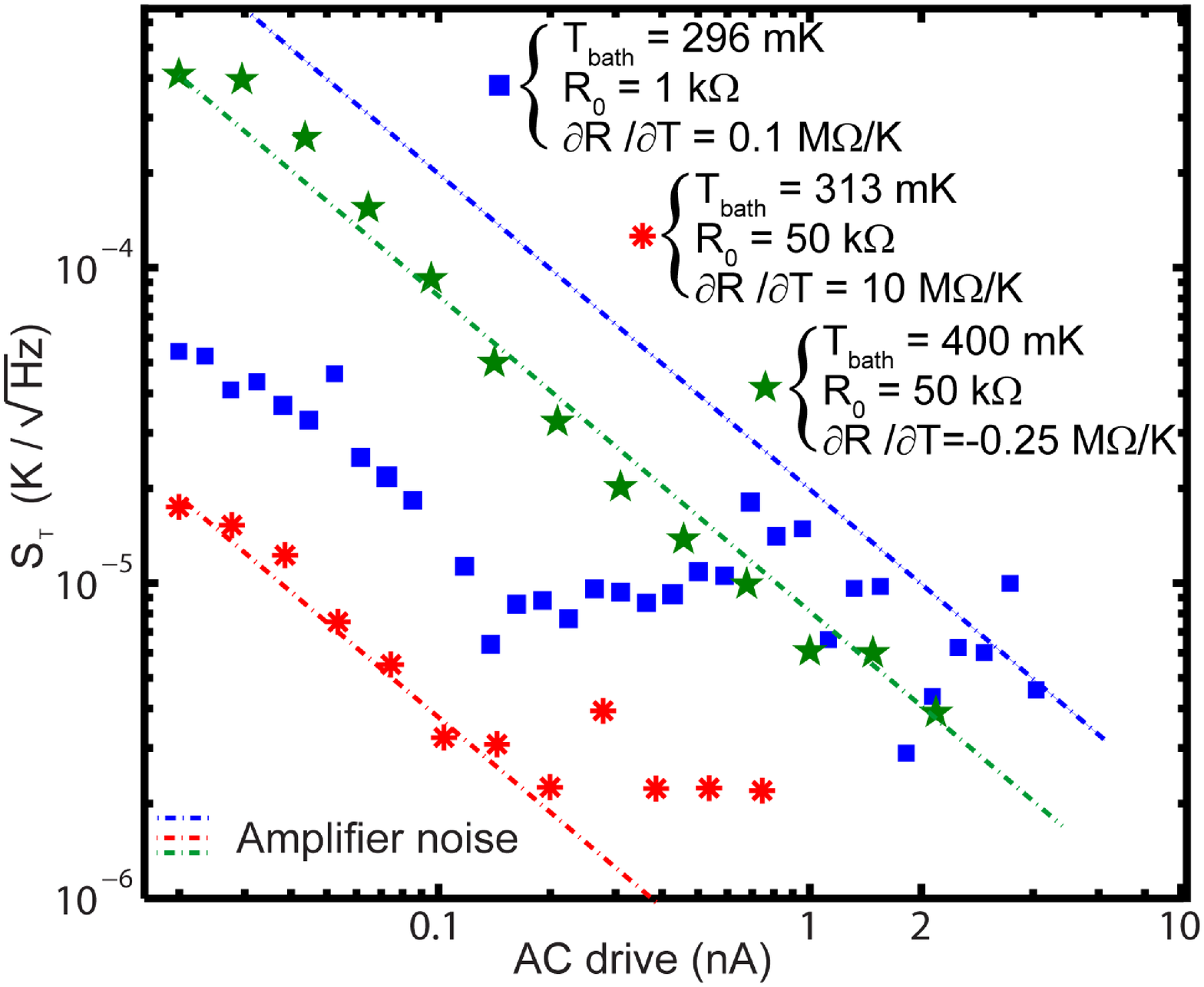}
\caption[ bath ]{Measured noise Equivalent Temperature (NET) of the SIS'IS structure is plotted for three different bath temperatures (color symbols).Straight lines are the fits to the Eq. (\ref{STexp}). For $T_{bath}$ = 296 mK, the low observed responsivity coupled to a low impedance lead to uncertainty that migh explain the low noise observed.  }
\label{fig:noise}
\end{figure}

Before measuring the noise, we have stabilized 
the bath temperature and measured the I-V curve of the device in order to find the switching 
current at this specific temperature. 
We have chosen the amplitude of the probe current to be half of the switching current value. 
Afterwards,  the bath temperature has been slowly modulated with the amplitude of less than 1 mK 
and with the frequency of 0.01 Hz, see Fig. \ref{fig:fft}b. 
We have monitored the time evolution of zero bias resistance (Fig. \ref{fig:fft}c) over several drive periods
and subsequently applied Fourier transformation to it. The result is plotted in Fig. \ref{fig:fft}a. 
A clear peak at 0.01 Hz corresponds to the frequency of the temperature drive, while
the plateau between 0.01 Hz and 0.1 Hz indicates the effective level of the temperature noise of the device, $S_T$,
which limits its resolution.
The bandwidth of the thermometer was determined by the lock-in time constant, which was rather long in our setup, $\tau_0\sim 300$ ms. 
In principle, this parameter can be safely reduced as long as it stays much longer than the inverse drive frequency.
The bandwidth of the thermometer can be increased to at least few kHz without loss of sensitivity, 
up to the cut-off frequency of the DC-lines.
  
In Fig. \ref{fig:noise} the dependence of the noise on the amplitude of the sinusoidal excitation current $I$ is plotted.
The temperature noise is the combination of amplifier noise and temperature noise\cite{Mather:82}, given by
\begin{eqnarray}
S_T = \sqrt{\left(\frac{S_V}{I\,\partial R_0/\partial T}\right)^2+\frac{2k_BT^2}{G_{th}}}.
\label{STexp}
\end{eqnarray}
In our device, the amplifier noise $S_V=4$ nV/$\sqrt{\rm Hz}$ dominates.
At the optimal operating temperature $T=313$ mK the noise saturates at the 
minimum value $S_T\approx 2\times 10^{-6}$ K/$\sqrt{\rm Hz}$
for probe currents higher than 0.2 nA, where $1/f$ background noise starts to dominate.

The performance of the radiation detector is usually characterized 
by noise equivalent power (NEP), which is estimated as
${\rm NEP}= G_{th}S_T$. For our device and in the absence of any THz radiation source, we find NEP $\approx 4\times 10^{-17}$ W/$\sqrt{\rm Hz}$. 
This value is about 10 times higher than that of the best conventional TES at 350 mK \cite{karasik_referee}, but it can be
reduced by optimization of the device parameters. First, one can use a better amplifier. For example, HEMT
amplifiers demonstrate the noise of 1 nV/$\sqrt{\rm Hz}$ at 4.2 K. 
Second, the heat conductance $G_{th}$ can be easily reduced, let say twice, by reducing the volume of the absorber.
A combination of these two simple measures already gives NEP$=5\times 10^{-18}$ W/$\sqrt{\rm Hz}$. 
More importantly, the critical temperature of titanium may be reduced 
down to 50 mK, or even below. In order to estimate NEP of our device at such a low temperature, 
we use Eqs. (\ref{Gth},\ref{eqn:zbslope})
and substitute the result in Eq. (\ref{STexp}). Keeping the same amplitude of
the driving current, i.e. $I=0.2$ nA, we arrive at the value ${\rm NEP}\sim 10^{-20}$ W/$\sqrt{\rm Hz}$, which
is comparable with what has been demonstrated for a traditional TES\cite{karasik_nature,karasik_referee}.
\section{Conclusion}
In summary, we have proposed a new type of a transition edge sensor, in which a superconducting
absorber is coupled to two superconducting leads by Josephson tunnel junctions.
We have fabricated the structure with titanium absorber and aluminum leads; it operates above 300 mK. 
By measuring its noise, we have demonstrated that the NEP of our device is comparable to that
of conventional TES sensors and can be improved further. However, the implementation of electro-thermal feedback\cite{Irwin} in our device is an important prerequisite to its practical use, since it would increase the dynamic range and make the working point more stable. This issue requires further research and we leave it beyond the scope of this paper. Our design allows one to tune the input impedance of the sensor in a wide range. It also ensures better confinement of exited quasiparticles in the absorber.  

This material is based upon work supported by the Academy of Finland under projects no. 139172 and 250280 (Center of Excellence), and by the European Commission under project no. 264034 ( Q-NET Marie Curie Initial Training Network). We would like to thank M.Meschke and J.Peltonen for the fruitful advices they gave on the fabrication of titanium tunnel junction. The research made use of the Otaniemi Research Infrastructure for Micro- and Nanotechnology  .


\end{document}